\documentclass[11pt,a4paper]{article}

\usepackage[utf8]{inputenc}
\usepackage[T1]{fontenc}
\usepackage{lmodern}
\usepackage{amsmath, amssymb}
\usepackage{geometry}
\usepackage{setspace}
\usepackage{xcolor}
\usepackage{listings}
\usepackage{booktabs}
\usepackage{algorithm}
\usepackage{algpseudocode}
\usepackage{hyperref}
\usepackage{float}
\usepackage{authblk}
\usepackage{microtype}

\definecolor{codegray}{gray}{0.95}
\definecolor{myblue}{RGB}{30,90,180}

\begin{document}

\title{\textbf{Eigen for Statistical and Machine Learning Computing: A Lightweight C++ Tutorial with Python Bindings}}

\author[1]{Seyoung Lee}
\author[2,3,*]{Kwan-Young Bak}

\affil[1]{\footnotesize Department of Statistics, Sungshin Women's University, Seoul 02844, South Korea, sudang0404@gmail.com}
\affil[2]{\footnotesize School of Mathematics, Statistics and Data Science, Sungshin Women's University, Seoul 02844, South Korea}
\affil[3]{\footnotesize Center for Data Science, Sungshin Women's University, Seoul 02844, South Korea}
\affil[*]{\footnotesize Corresponding author: kybak@sungshin.ac.kr}

\date{}
\maketitle

\begin{abstract}
This note provides a lightweight tutorial on using Eigen, a C++ template library for linear algebra, to implement statistical and machine learning algorithms. The emphasis is practical rather than methodological: we show how common matrix operations, decomposition-based solvers, and vectorized updates can be written in readable C++ and then connected to Python through \texttt{pybind11}. Two examples are used throughout the tutorial: kernel ridge regression and matrix factorization with stochastic gradient descent. The examples are intentionally small enough to be studied as code, but they contain many operations that appear in larger research software projects, including kernel matrix construction, regularized linear system solving, row-wise updates, and NumPy--Eigen data conversion. The goal is to provide a reproducible starting point for researchers who want to move from mathematical formulas to efficient C++ implementations while retaining a convenient Python workflow.
\end{abstract}

\section{Introduction}

Many statistical and machine learning algorithms can be expressed as sequences of linear algebra operations: constructing design or kernel matrices, computing inner products, solving regularized linear systems, applying matrix decompositions, or updating parameter vectors iteratively. These operations are often hidden behind high-level software, but they become central when researchers need to implement new estimators, modify existing algorithms, or build reusable computational backends.

Python and R are convenient environments for data analysis, experimentation, and visualization. Python is especially common in machine learning workflows through tools such as NumPy and scikit-learn \cite{harris2020array,pedregosa2011scikit}. However, when the computational core of a method is new or performance-sensitive, it can be useful to implement that core in C++ and expose it to a high-level language. This backend/frontend separation allows the expensive numerical computation to be written in a compiled language while preserving an accessible interface for applied users.

Eigen is well suited to this role because it is a header-only C++ library with expressive matrix and vector syntax \cite{eigenweb}. It supports dense and sparse matrices, fixed-size and dynamic-size objects, matrix decompositions, numerical solvers, and expression-template-based optimizations. For researchers who already think in matrix notation, Eigen often provides a direct route from a mathematical formula to C++ code. It can also be linked with optimized BLAS/LAPACK backends when large dense matrix computations dominate the runtime \cite{blas1979,lapack1999}.

This paper is intentionally written as a compact tutorial note rather than as a broad review of numerical linear algebra libraries or research software ecosystems. The contribution of the note is threefold. First, it provides a minimal but complete Eigen-based implementation pattern for matrix-based statistical computing. Second, it illustrates how mathematical estimators can be translated into reusable C++ classes through two representative examples: kernel ridge regression and matrix factorization. Third, it shows how such C++ implementations can be exposed to Python through \texttt{pybind11} \cite{pybind11} while preserving a familiar NumPy-based workflow. The scope of this note is deliberately limited to Python bindings. Interfaces to other high-level languages, including R, raise additional questions about package structure, memory ownership, and distribution workflows; these issues are therefore beyond the scope of this note. All source code for the Eigen-based C++ implementations, Python bindings, and benchmark examples is available in the accompanying GitHub repository: \url{https://github.com/seyoung-230/EigenpyR}.

\section{Minimal Eigen Operations Used in the Examples}

This section summarizes the Eigen operations used later in the tutorial. The aim is not to cover the full Eigen API, but to establish a small working vocabulary for matrix-based statistical computing.

\subsection{Matrices and Vectors}

Dynamically sized vectors and matrices are represented by \texttt{VectorXd} and \texttt{MatrixXd}:

\begin{lstlisting}
VectorXd x(n);
MatrixXd X(n, p);
\end{lstlisting}

These objects correspond directly to common statistical quantities such as response vectors, coefficient vectors, design matrices, kernel matrices, and latent factor matrices. When dimensions are known at compile time, fixed-size types such as \texttt{Vector3d} or \texttt{Matrix<double, 3, 3>} can also be used.

\subsection{Initialization and Arithmetic}

Eigen supports concise initialization and arithmetic syntax:

\begin{lstlisting}
VectorXd v(3);
v << 1, 2, 3;

MatrixXd A(2, 2);
A << 1, 2,
     3, 4;

MatrixXd I = MatrixXd::Identity(3, 3);
MatrixXd Z = MatrixXd::Zero(3, 3);
MatrixXd C = A + B;
MatrixXd D = A * B;
VectorXd y = A * x;
\end{lstlisting}

For matrix objects, \texttt{*} denotes matrix multiplication. Element-wise operations are accessed through the array interface:

\begin{lstlisting}
A.array() * B.array();
A.array().exp();
A.array().log();
\end{lstlisting}

This distinction is useful because it makes the difference between linear algebra operations and component-wise operations explicit.

\subsection{Rows, Columns, Blocks, and Inner Products}

Many algorithms operate on rows, columns, or submatrices. Eigen provides direct access to these objects:

\begin{lstlisting}
A(i, j);             // element access
A.row(i);            // i-th row
A.col(j);            // j-th column
A.block(i, j, r, c); // r by c block starting at (i, j)
\end{lstlisting}

Inner products, norms, and transposes are also directly available:

\begin{lstlisting}
double d = a.dot(b);
double norm = a.norm();
MatrixXd At = A.transpose();
\end{lstlisting}

These operations appear below in the Gaussian kernel computation and in the row-wise updates for matrix factorization.

\subsection{Solving Linear Systems}

Many estimators require solving a system of the form \(Ax=b\). In numerical computing, it is usually preferable to solve such systems by matrix factorization rather than by explicitly computing \(A^{-1}\). Eigen provides several decomposition-based solvers. For example,

\begin{lstlisting}
VectorXd x = A.colPivHouseholderQr().solve(b);
VectorXd z = A.ldlt().solve(b);
\end{lstlisting}

The second form is useful for symmetric regularized systems such as those appearing in ridge regression and kernel ridge regression. This pattern is one of the main reasons why Eigen is convenient for statistical algorithms: the implementation can remain close to the mathematical expression while following standard numerical practice.

\subsection{Performance Notes}

Eigen is header-only and can be used without linking to external numerical libraries. It also uses expression templates and lazy evaluation to reduce unnecessary temporary objects. When assigning a matrix product to a destination that does not alias either operand, \texttt{.noalias()} can be used:

\begin{lstlisting}
C.noalias() = A * B;
\end{lstlisting}

For larger dense computations, Eigen can optionally use BLAS/LAPACK backends. A small benchmark for the kernel ridge regression example is included in Appendix~\ref{app:blas-benchmark}.

\section{Example 1: Kernel Ridge Regression}

Kernel ridge regression (KRR) combines least-squares estimation with kernel methods \cite{saunders1998ridge,scholkopf2002learning,shaweTaylor2004kernel}. It is a useful tutorial example because fitting reduces to kernel matrix construction and the solution of a regularized linear system.

We use the Gaussian radial basis function kernel
\[
K(x_i,x_j)=\exp\left(-\frac{\|x_i-x_j\|^2}{2\sigma^2}\right),
\]
where \(\sigma>0\) is the bandwidth parameter. Given training data \((X,y)\), the response vector is centered as
\[
y_c = y - \bar y.
\]
The dual coefficient vector \(\alpha\) is obtained by solving
\[
(K+\lambda I)\alpha = y_c,
\]
where \(\lambda\geq 0\) is the regularization parameter. Prediction for new inputs \(X_{\mathrm{new}}\) is
\[
\hat y = K(X_{\mathrm{new}},X)\alpha + \bar y.
\]

\begin{algorithm}[H]
{\footnotesize
\caption{Kernel Ridge Regression}
\begin{algorithmic}[1]
\State \textbf{Input:} training matrix $X$, response vector $y$, regularization parameter $\lambda$, bandwidth $\sigma$
\State Compute $\bar y$ and set $y_c \gets y-\bar y$
\State Compute the kernel matrix $K$ with entries $K(x_i,x_j)$
\State Solve $(K+\lambda I)\alpha = y_c$
\Statex
\State \textbf{Prediction:}
\State Compute $K(X_{\mathrm{new}},X)$
\State $\hat y \gets K(X_{\mathrm{new}},X)\alpha + \bar y$
\State \Return $\hat y$
\end{algorithmic}
}
\end{algorithm}

In the C++ implementation, the Gaussian kernel matrix is constructed from pairwise squared distances. For two input matrices \(A\) and \(B\), one entry can be computed as

\begin{lstlisting}
double dist2 = (A.row(i) - B.row(j)).squaredNorm();
K(i, j) = std::exp(-dist2 / (2.0 * sigma_ * sigma_));
\end{lstlisting}

The response is centered using Eigen's array interface:

\begin{lstlisting}
VectorXd y_centered = y.array() - y_mean_;
\end{lstlisting}

The regularized system is solved using \texttt{LDLT} decomposition:

\begin{lstlisting}
alpha_ = (K + lambda_ * I).ldlt().solve(y_centered);
\end{lstlisting}

Prediction is then a matrix--vector multiplication followed by adding the response mean:

\begin{lstlisting}
VectorXd pred = K_new * alpha_;
pred.array() += y_mean_;
\end{lstlisting}

The complete C++ implementation is provided in Appendix~\ref{app:krr-code}.

\section{Example 2: Matrix Factorization with Stochastic Gradient Descent}

Matrix factorization is widely used in recommendation systems and latent representation learning \cite{koren2009matrix}. It is also a useful Eigen example because it uses sparse observations, row-wise matrix access, inner products, random initialization, and iterative updates.

For an observed rating \((u,i,r_{ui})\), the model predicts
\[
\hat r_{ui}=\mu+b_u+b_i+p_u^\top q_i,
\]
where \(\mu\) is the global mean, \(b_u\) and \(b_i\) are user and item biases, and \(p_u,q_i\in\mathbb R^k\) are latent factor vectors. A common objective is
\[
\sum_{(u,i)\in\Omega}(r_{ui}-\hat r_{ui})^2
+\lambda\left(\|p_u\|^2+\|q_i\|^2+b_u^2+b_i^2\right).
\]

\begin{algorithm}[H]
{\footnotesize
\caption{Matrix Factorization via SGD}
\begin{algorithmic}[1]
\State \textbf{Input:} observed ratings $\Omega$, latent dimension $k$, learning rate $\eta$, regularization parameter $\lambda$, number of epochs $T$
\State Initialize $P$ and $Q$ from a normal distribution
\State Initialize $b_u=0$ and $b_i=0$
\State Compute global mean $\mu$
\For{epoch $=1,\dots,T$}
    \State Shuffle $\Omega$
    \For{each $(u,i,r_{ui})\in\Omega$}
        \State $\hat r_{ui}\gets \mu+b_u+b_i+p_u^\top q_i$
        \State $e_{ui}\gets r_{ui}-\hat r_{ui}$
        \State $b_u \gets b_u + \eta(e_{ui}-\lambda b_u)$
        \State $b_i \gets b_i + \eta(e_{ui}-\lambda b_i)$
        \State $p_u \gets p_u + \eta(e_{ui}q_i-\lambda p_u)$
        \State $q_i \gets q_i + \eta(e_{ui}p_u-\lambda q_i)$
    \EndFor
\EndFor
\State \Return $P,Q,b_u,b_i,\mu$
\end{algorithmic}
}
\end{algorithm}

Observed ratings are stored as a list of triplets using \texttt{std::vector<Rating>} rather than as a dense user--item matrix. The predicted rating is computed by an Eigen inner product:

\begin{lstlisting}
double pred = mu_ + bu_(u) + bi_(i) + P_.row(u).dot(Q_.row(i));
\end{lstlisting}

Before updating the two factor vectors, the current row values are copied:

\begin{lstlisting}
Eigen::RowVectorXd pu = P_.row(u);
Eigen::RowVectorXd qi = Q_.row(i);
\end{lstlisting}

The SGD updates are then implemented as

\begin{lstlisting}
bu_(u) += lr_ * (err - reg_ * bu_(u));
bi_(i) += lr_ * (err - reg_ * bi_(i));

P_.row(u) += lr_ * (err * qi - reg_ * pu);
Q_.row(i) += lr_ * (err * pu - reg_ * qi);
\end{lstlisting}

The complete implementation is provided in Appendix~\ref{app:mfsgd-code}.

\section{Python Bindings with \texttt{pybind11}}

The previous examples define ordinary C++ classes. To use them from Python, we expose the classes through \texttt{pybind11}. This allows the computationally intensive routines to remain in C++ while the user-facing workflow remains close to ordinary Python code.

The extension module is defined using the \texttt{PYBIND11\_MODULE} macro and can be imported as

\begin{lstlisting}[language=Python]
import eigenpyr
\end{lstlisting}

The header \texttt{pybind11/eigen.h} enables automatic conversion between Eigen matrices/vectors and NumPy arrays. The header \texttt{pybind11/stl.h} supports conversions between C++ standard library containers and Python objects.

\subsection{Binding Kernel Ridge Regression}

The \texttt{KernelRidge} class is exposed as \texttt{eigenpyr.KernelRidge}:

\begin{lstlisting}
static void bind_kernel_ridge(py::module_& m) {
    py::class_<KernelRidge>(m, "KernelRidge")
        .def(py::init<double, double>(),
             py::arg("lambda_"),
             py::arg("sigma"))
        .def("fit", &KernelRidge::fit,
             py::arg("X"),
             py::arg("y"))
        .def("predict", &KernelRidge::predict,
             py::arg("X_new"))
        .def_property_readonly("alpha", &KernelRidge::alpha)
        .def_property_readonly("x_train", &KernelRidge::x_train)
        .def_property_readonly("lambda_", &KernelRidge::lambda)
        .def_property_readonly("sigma", &KernelRidge::sigma)
        .def_property_readonly("y_mean", &KernelRidge::y_mean);
}
\end{lstlisting}

A minimal Python example is

\begin{lstlisting}[language=Python]
import numpy as np
import eigenpyr

X = np.linspace(-1, 1, 100).reshape(-1, 1)
y = np.sin(2 * np.pi * X[:, 0])

model = eigenpyr.KernelRidge(lambda_=0.001, sigma=0.2)
model.fit(X, y)

X_new = np.linspace(-1, 1, 20).reshape(-1, 1)
pred = model.predict(X_new)
\end{lstlisting}

From the Python user's perspective, \texttt{X} and \texttt{y} are ordinary NumPy arrays. Internally, they are converted to Eigen objects and passed to the C++ implementation.

\subsection{Binding Matrix Factorization}

For matrix factorization, the \texttt{Rating} structure is exposed first:

\begin{lstlisting}
py::class_<Rating>(m, "Rating")
    .def(py::init<>())
    .def_readwrite("user", &Rating::user)
    .def_readwrite("item", &Rating::item)
    .def_readwrite("value", &Rating::value);
\end{lstlisting}

The main model class is exposed as \texttt{eigenpyr.MatrixFactorizationSGD}:

\begin{lstlisting}
py::class_<MatrixFactorizationSGD>(m, "MatrixFactorizationSGD")
    .def(
        py::init<int, int, int, double, double, int, unsigned int>(),
        py::arg("n_users"),
        py::arg("n_items"),
        py::arg("n_factors") = 10,
        py::arg("lr") = 0.01,
        py::arg("reg") = 0.02,
        py::arg("n_epochs") = 20,
        py::arg("seed") = 42
    )
    .def("fit", &MatrixFactorizationSGD::fit,
         py::arg("ratings"),
         py::arg("verbose") = true)
    .def("predict", &MatrixFactorizationSGD::predict,
         py::arg("user"),
         py::arg("item"))
    .def("full_prediction", &MatrixFactorizationSGD::full_prediction)
    .def_property_readonly("user_factors", &MatrixFactorizationSGD::user_factors)
    .def_property_readonly("item_factors", &MatrixFactorizationSGD::item_factors)
    .def_property_readonly("user_bias", &MatrixFactorizationSGD::user_bias)
    .def_property_readonly("item_bias", &MatrixFactorizationSGD::item_bias)
    .def_property_readonly("global_mean", &MatrixFactorizationSGD::global_mean);
\end{lstlisting}

A minimal Python workflow is

\begin{lstlisting}[language=Python]
import eigenpyr

ratings = []

r1 = eigenpyr.Rating()
r1.user = 0
r1.item = 0
r1.value = 5.0
ratings.append(r1)

r2 = eigenpyr.Rating()
r2.user = 0
r2.item = 1
r2.value = 3.0
ratings.append(r2)

r3 = eigenpyr.Rating()
r3.user = 1
r3.item = 0
r3.value = 4.0
ratings.append(r3)

model = eigenpyr.MatrixFactorizationSGD(
    n_users=2,
    n_items=2,
    n_factors=5,
    lr=0.01,
    reg=0.02,
    n_epochs=20,
    seed=42
)

model.fit(ratings, verbose=True)
pred = model.predict(0, 1)
full_pred = model.full_prediction()
\end{lstlisting}

This example shows how sparse C++ data structures and Eigen objects can be used internally while the Python interface remains simple.

\section{Summary}

This tutorial note presented a minimal Eigen-based workflow for implementing statistical and machine learning examples in C++ and exposing them to Python. The KRR example illustrated kernel matrix construction and regularized linear system solving. The matrix factorization example illustrated sparse rating data, row-wise matrix operations, and iterative SGD updates. The binding examples showed how \texttt{pybind11} connects C++ classes with NumPy-compatible Python workflows.

\newpage
\bibliographystyle{plain}
\bibliography{eigen_arxiv_light_references}

\newpage
\appendix
\section{Installing Eigen and CMake Setup}
\label{app:eigen-install}

Eigen is a header-only C++ library. Therefore, no separate compilation of
Eigen itself is required. The main requirement is that the compiler and CMake
can locate the Eigen header files. The commands below are intended as minimal
examples for common platforms. Exact package names, compiler paths, and CMake
options may vary depending on the local environment. The accompanying
repository should be consulted for the most up-to-date build instructions.

\subsection{Basic Idea}

After installation, Eigen is typically included in C++ source files as

\begin{lstlisting}
#include <Eigen/Dense>
\end{lstlisting}

For CMake-based projects, Eigen is usually linked as an interface target:

\begin{lstlisting}
find_package(Eigen3 REQUIRED)
target_link_libraries(my_target PRIVATE Eigen3::Eigen)
\end{lstlisting}

Since Eigen is header-only, this does not link to a compiled Eigen library.
Instead, it tells the compiler where to find the Eigen headers.

\subsection{Windows}

\subsubsection{Using vcpkg}

One convenient way to install Eigen on Windows is through \texttt{vcpkg}:

\begin{lstlisting}
vcpkg install eigen3:x64-windows
\end{lstlisting}

A minimal CMake configuration for a standalone executable is

\begin{lstlisting}
cmake_minimum_required(VERSION 3.15)
project(EigenExample)

set(CMAKE_CXX_STANDARD 17)
set(CMAKE_CXX_STANDARD_REQUIRED ON)

find_package(Eigen3 CONFIG REQUIRED)

add_executable(eigen_example src/main.cpp)
target_link_libraries(eigen_example PRIVATE Eigen3::Eigen)
\end{lstlisting}

The project can then be configured and built by specifying the vcpkg toolchain
file. The exact path depends on where \texttt{vcpkg} is installed.

\begin{lstlisting}
cmake -B build -S . ^
  -DCMAKE_TOOLCHAIN_FILE=<path-to-vcpkg>/scripts/buildsystems/vcpkg.cmake

cmake --build build --config Release
\end{lstlisting}

If an environment variable such as \texttt{VCPKG\_ROOT} is defined, the
toolchain file can also be specified as

\begin{lstlisting}
cmake -B build -S . ^
  -DCMAKE_TOOLCHAIN_FILE=%VCPKG_ROOT%/scripts/buildsystems/vcpkg.cmake
\end{lstlisting}

\subsubsection{Manual installation}

Eigen can also be installed manually by downloading and extracting the Eigen
source directory. Suppose Eigen is extracted to

\begin{lstlisting}
C:/Libraries/eigen
\end{lstlisting}

Then the file \texttt{Eigen/Dense} should be located at

\begin{lstlisting}
C:/Libraries/eigen/Eigen/Dense
\end{lstlisting}

In this case, the include directory can be added manually:

\begin{lstlisting}
cmake_minimum_required(VERSION 3.15)
project(EigenExample)

set(CMAKE_CXX_STANDARD 17)
set(CMAKE_CXX_STANDARD_REQUIRED ON)

add_executable(eigen_example src/main.cpp)

target_include_directories(eigen_example PRIVATE
    C:/Libraries/eigen
)
\end{lstlisting}

\subsection{macOS}

On macOS, Eigen can be installed using Homebrew:

\begin{lstlisting}
brew install eigen
\end{lstlisting}

A minimal CMake configuration is

\begin{lstlisting}
cmake_minimum_required(VERSION 3.15)
project(EigenExample)

set(CMAKE_CXX_STANDARD 17)
set(CMAKE_CXX_STANDARD_REQUIRED ON)

find_package(Eigen3 REQUIRED)

add_executable(eigen_example src/main.cpp)
target_link_libraries(eigen_example PRIVATE Eigen3::Eigen)
\end{lstlisting}

The project can usually be built as follows:

\begin{lstlisting}
cmake -B build -S .
cmake --build build
\end{lstlisting}

If CMake cannot find Eigen automatically, the Homebrew installation path can be
provided through \texttt{CMAKE\_PREFIX\_PATH}:

\begin{lstlisting}
cmake -B build -S . -DCMAKE_PREFIX_PATH="$(brew --prefix eigen)"
cmake --build build
\end{lstlisting}

\subsection{Linux}

On Ubuntu or Debian-based systems, Eigen can be installed through \texttt{apt}:

\begin{lstlisting}
sudo apt update
sudo apt install libeigen3-dev
\end{lstlisting}

A minimal CMake configuration is

\begin{lstlisting}
cmake_minimum_required(VERSION 3.15)
project(EigenExample)

set(CMAKE_CXX_STANDARD 17)
set(CMAKE_CXX_STANDARD_REQUIRED ON)

find_package(Eigen3 REQUIRED)

add_executable(eigen_example src/main.cpp)
target_link_libraries(eigen_example PRIVATE Eigen3::Eigen)
\end{lstlisting}

The project can then be configured and built by

\begin{lstlisting}
cmake -B build -S .
cmake --build build
\end{lstlisting}

\subsection{Minimal Standalone CMake Template}

For a simple Eigen-based C++ executable, the following \texttt{CMakeLists.txt}
is sufficient on most systems where Eigen is discoverable by CMake:

\begin{lstlisting}
cmake_minimum_required(VERSION 3.15)
project(EigenExample)

set(CMAKE_CXX_STANDARD 17)
set(CMAKE_CXX_STANDARD_REQUIRED ON)

find_package(Eigen3 REQUIRED)

add_executable(eigen_example
    src/main.cpp
)

target_link_libraries(eigen_example PRIVATE Eigen3::Eigen)
\end{lstlisting}

A minimal source file may look as follows:

\begin{lstlisting}
#include <Eigen/Dense>
#include <iostream>

int main() {
    Eigen::MatrixXd A(2, 2);
    A << 1.0, 2.0,
         3.0, 4.0;

    Eigen::VectorXd x(2);
    x << 1.0, 1.0;

    Eigen::VectorXd y = A * x;

    std::cout << y << std::endl;
    return 0;
}
\end{lstlisting}

\subsection{Minimal CMake Template for a Python Module}

When Eigen-based C++ code is exposed to Python through \texttt{pybind11}, the
build target is a Python extension module rather than a standalone executable.
This configuration assumes that \texttt{pybind11} is already installed and
discoverable by CMake. A minimal configuration is

\begin{lstlisting}
cmake_minimum_required(VERSION 3.15)
project(EigenPyR)

set(CMAKE_CXX_STANDARD 17)
set(CMAKE_CXX_STANDARD_REQUIRED ON)

find_package(Eigen3 REQUIRED)
find_package(pybind11 CONFIG REQUIRED)

pybind11_add_module(eigenpyr
    src/bindings.cpp
    src/kernel_ridge.cpp
    src/mf_sgd.cpp
)

target_link_libraries(eigenpyr PRIVATE Eigen3::Eigen)
\end{lstlisting}

Additional include directories may be required depending on where the project
header files are located.

In the binding source file, Eigen and NumPy conversion support is enabled by
including \texttt{pybind11/eigen.h}:

\begin{lstlisting}
#include <pybind11/pybind11.h>
#include <pybind11/eigen.h>
#include <pybind11/stl.h>

namespace py = pybind11;
\end{lstlisting}

The module can then be built using CMake:

\begin{lstlisting}
cmake -B build -S .
cmake --build build
\end{lstlisting}

After compilation, the generated Python extension module can be imported in
Python, provided that the compiled module is located in the current working
directory or on the Python module search path:

\begin{lstlisting}
import eigenpyr
\end{lstlisting}

\subsection{Common Build Notes}

For numerical experiments and benchmarks, the project should be compiled in
Release mode:

\begin{lstlisting}
cmake -B build -S . -DCMAKE_BUILD_TYPE=Release
cmake --build build
\end{lstlisting}

On multi-configuration generators such as Visual Studio, the configuration is
usually selected at build time:

\begin{lstlisting}
cmake --build build --config Release
\end{lstlisting}

If Eigen is installed but CMake cannot locate it, one can provide the
installation prefix manually:

\begin{lstlisting}
cmake -B build -S . -DCMAKE_PREFIX_PATH=<path-to-installed-eigen>
\end{lstlisting}

This option is mainly useful when Eigen was installed through a package manager
or another installation procedure that provides CMake configuration files.

If Eigen is used manually without \texttt{find\_package}, the include directory
should be the directory that contains the \texttt{Eigen} folder. For example,
if the file \texttt{Eigen/Dense} is located at

\begin{lstlisting}
/home/user/libraries/eigen/Eigen/Dense
\end{lstlisting}

then the include directory should be

\begin{lstlisting}
/home/user/libraries/eigen
\end{lstlisting}

\section{BLAS Integration and Benchmark Results}
\label{app:blas-benchmark}

This appendix briefly describes how Eigen can be linked with an external BLAS backend and reports benchmark results for the kernel ridge regression implementation.

\subsection{BLAS Integration}

Eigen can be used without external numerical libraries because it provides its own internal computational engine. However, Eigen also supports optional integration with external BLAS libraries for selected dense matrix operations. In this work, the BLAS-linked version was compiled by defining the \texttt{EIGEN\_USE\_BLAS} macro and linking the executable to OpenBLAS through CMake.

A simplified CMake configuration is shown below.

\begin{lstlisting}
target_compile_definitions(bench_krr_blas PRIVATE EIGEN_USE_BLAS)
target_link_libraries(bench_krr_blas PRIVATE openblas)
\end{lstlisting}

The executable was built in Release mode. The default Eigen version was compiled without \texttt{EIGEN\_USE\_BLAS}, so Eigen's internal computational engine was used. The exact library name and linking options may differ depending on the
operating system, compiler, and BLAS installation method.

\subsection{Benchmark Setting}

The benchmark compares kernel ridge regression implemented using Eigen's internal engine, Eigen linked with BLAS, and Armadillo~\cite{sanderson2016armadillo}. The sample size was set to \(n = 1000, 2000, 3000, 4000\), with input dimension \(d = 5\) and \(n_{\mathrm{test}} = 200\). The regularization parameter and Gaussian kernel bandwidth were fixed at \(\lambda = 10^{-4}\) and \(\sigma = 1\). Each experiment was repeated 15 times, and the average fitting times were reported.

The experiments were conducted under the following environment.

\begin{table}[H]
\centering
\begin{tabular}{ll}
\toprule
Component & Specification \\
\midrule
Operating system & Windows 11 64-bit \\
CPU & Intel Core i7-14700 \\
Memory & 32 GB RAM \\
Compiler & g++ 15.2.0 \\
Build system & CMake \\
CMake version & 4.1.0-rc1 \\
Eigen version & 5.0.0 \\
Armadillo version & 14.6.0 \\
BLAS backend & OpenBLAS \\
Build type & Release \\
Number of runs & 15 \\
\bottomrule
\end{tabular}
\caption{Computing environment used for the benchmark experiments.}
\label{tab:benchmark-environment}
\end{table}

\subsection{Benchmark Results}

\begin{table}[H]
\centering
\begin{tabular}{llc}
\toprule
\(n\) & Method & Fit Time (s) \\
\midrule
1000 & Eigen & 0.0249356 \\
1000 & Armadillo & 0.0200975 \\
\midrule
2000 & Eigen & 0.157169 \\
2000 & Armadillo & 0.0756702 \\
\midrule
3000 & Eigen & 0.481685 \\
3000 & Armadillo & 0.174085 \\
\midrule
4000 & Eigen & 1.09485 \\
4000 & Armadillo & 0.340826 \\
\bottomrule
\end{tabular}
\caption{KRR fitting benchmark results for Eigen with its internal computational engine and Armadillo. Results are averaged over 15 runs.}
\label{tab:krr-eigen-core-armadillo}
\end{table}

\begin{table}[H]
\centering
\begin{tabular}{llc}
\toprule
\(n\) & Method & Fit Time (s) \\
\midrule
1000 & Eigen + BLAS & 0.0156935 \\
1000 & Armadillo & 0.0182886 \\
\midrule
2000 & Eigen + BLAS & 0.0578315 \\
2000 & Armadillo & 0.0731369 \\
\midrule
3000 & Eigen + BLAS & 0.133951 \\
3000 & Armadillo & 0.169586 \\
\midrule
4000 & Eigen + BLAS & 0.257884 \\
4000 & Armadillo & 0.330168 \\
\bottomrule
\end{tabular}
\caption{KRR fitting benchmark results for Eigen linked with BLAS and Armadillo. Results are averaged over 15 runs.}
\label{tab:krr-eigen-blas-armadillo}
\end{table}

In this benchmark environment, the BLAS-linked Eigen implementation substantially reduced fitting time compared with the default Eigen implementation. For example, when \(n=4000\), the fitting time decreased from 1.09485 seconds to 0.257884 seconds after BLAS integration. The BLAS-linked Eigen implementation was also faster than Armadillo in the tested settings. These results should be interpreted as environment-specific rather than as a general performance ranking. Overall, the benchmark suggests that Eigen's default internal engine is convenient for portability, while BLAS integration can be useful when dense matrix computations dominate the runtime.

\section{Complete Implementation of Kernel Ridge Regression}
\label{app:krr-code}

This appendix provides the full C++ implementation of the \texttt{KernelRidge} class.

\subsection{Header File}

\begin{lstlisting}
#ifndef SES_KERNEL_RIDGE_H
#define SES_KERNEL_RIDGE_H

#include <Eigen/Dense>
using namespace Eigen;

class KernelRidge {
public:
    KernelRidge(double lambda, double sigma);

    void fit(const MatrixXd& X, const VectorXd& y);
    VectorXd predict(const MatrixXd& X_new) const;

    const VectorXd& alpha() const { return alpha_; }
    const MatrixXd& x_train() const { return X_train_; }
    double lambda() const { return lambda_; }
    double sigma() const { return sigma_; }
    double y_mean() const { return y_mean_; }

private:
    MatrixXd kernel_matrix(const MatrixXd& A, const MatrixXd& B) const;

    MatrixXd X_train_;
    VectorXd alpha_;
    double lambda_ = 0.0;
    double sigma_ = 1.0;
    double y_mean_ = 0.0;
};

#endif // SES_KERNEL_RIDGE_H
\end{lstlisting}

\subsection{Source File}

\begin{lstlisting}
#include "kernel_ridge.h"

#include <cmath>
#include <stdexcept>

KernelRidge::KernelRidge(double lambda, double sigma)
    : lambda_(lambda), sigma_(sigma) {
    if (lambda_ < 0.0) {
        throw std::invalid_argument("lambda must be non-negative.");
    }
    if (sigma_ <= 0.0) {
        throw std::invalid_argument("sigma must be positive.");
    }
}

MatrixXd KernelRidge::kernel_matrix(const MatrixXd& A,
                                    const MatrixXd& B) const {
    if (A.cols() != B.cols()) {
        throw std::invalid_argument(
            "A and B must have the same number of columns."
        );
    }

    int n1 = static_cast<int>(A.rows());
    int n2 = static_cast<int>(B.rows());
    MatrixXd K(n1, n2);

    for (int i = 0; i < n1; ++i) {
        for (int j = 0; j < n2; ++j) {
            double dist2 =
                (A.row(i) - B.row(j)).squaredNorm();
            K(i, j) =
                std::exp(-dist2 / (2.0 * sigma_ * sigma_));
        }
    }

    return K;
}

void KernelRidge::fit(const MatrixXd& X,
                      const VectorXd& y) {
    if (X.rows() != y.size()) {
        throw std::invalid_argument(
            "Number of rows in X must match length of y."
        );
    }

    X_train_ = X;

    y_mean_ = y.mean();
    VectorXd y_centered =
        y.array() - y_mean_;

    int n = static_cast<int>(X.rows());
    MatrixXd K =
        kernel_matrix(X_train_, X_train_);
    MatrixXd I =
        MatrixXd::Identity(n, n);

    alpha_ =
        (K + lambda_ * I)
        .ldlt()
        .solve(y_centered);
}

VectorXd KernelRidge::predict(
    const MatrixXd& X_new) const {

    if (X_train_.rows() == 0 ||
        alpha_.size() == 0) {
        throw std::runtime_error(
            "Model has not been fitted yet."
        );
    }

    MatrixXd K_new =
        kernel_matrix(X_new, X_train_);

    VectorXd pred = K_new * alpha_;
    pred.array() += y_mean_;

    return pred;
}
\end{lstlisting}

\section{Complete Implementation of Matrix Factorization with SGD}
\label{app:mfsgd-code}

This appendix provides the full C++ implementation of the
\texttt{MatrixFactorizationSGD} class.

\subsection{Header File}

\begin{lstlisting}
#ifndef SES_MF_SGD_H
#define SES_MF_SGD_H

#include <Eigen/Dense>
#include <vector>
#include <random>
#include <stdexcept>
using namespace Eigen;

struct Rating {
    int user;
    int item;
    double value;
};

class MatrixFactorizationSGD {
public:
    MatrixFactorizationSGD(
        int n_users,
        int n_items,
        int n_factors = 10,
        double lr = 0.01,
        double reg = 0.02,
        int n_epochs = 20,
        unsigned int seed = 42
    );

    void fit(const std::vector<Rating>& ratings, bool verbose = true);

    double predict(int user, int item) const;
    MatrixXd full_prediction() const;

    const MatrixXd& user_factors() const;
    const MatrixXd& item_factors() const;
    const VectorXd& user_bias() const;
    const VectorXd& item_bias() const;
    double global_mean() const;

private:
    int n_users_;
    int n_items_;
    int n_factors_;
    double lr_;
    double reg_;
    int n_epochs_;
    unsigned int seed_;
    double global_mean_;

    MatrixXd P_;
    MatrixXd Q_;
    VectorXd bu_;
    VectorXd bi_;

    std::mt19937 rng_;

    void initialize_parameters();
    void validate_indices(const std::vector<Rating>& ratings) const;
    double compute_rmse(const std::vector<Rating>& ratings) const;
};

#endif //SES_MF_SGD_H
\end{lstlisting}

\subsection{Source File}

\begin{lstlisting}
#include "mf_sgd.h"

#include <algorithm>
#include <iostream>
#include <cmath>

MatrixFactorizationSGD::MatrixFactorizationSGD(
    int n_users,
    int n_items,
    int n_factors,
    double lr,
    double reg,
    int n_epochs,
    unsigned int seed
)
    : n_users_(n_users),
      n_items_(n_items),
      n_factors_(n_factors),
      lr_(lr),
      reg_(reg),
      n_epochs_(n_epochs),
      seed_(seed),
      global_mean_(0.0),
      P_(n_users, n_factors),
      Q_(n_items, n_factors),
      bu_(n_users),
      bi_(n_items),
      rng_(seed) {

    if (n_users_ <= 0 || n_items_ <= 0 || n_factors_ <= 0) {
        throw std::invalid_argument("n_users, n_items, n_factors must be positive.");
    }
    if (lr_ <= 0.0) {
        throw std::invalid_argument("learning rate must be positive.");
    }
    if (reg_ < 0.0) {
        throw std::invalid_argument("regularization must be non-negative.");
    }
    if (n_epochs_ <= 0) {
        throw std::invalid_argument("n_epochs must be positive.");
    }

    initialize_parameters();
}

void MatrixFactorizationSGD::initialize_parameters() {
    std::normal_distribution<double> dist(0.0, 0.1);

    for (int u = 0; u < n_users_; ++u) {
        for (int k = 0; k < n_factors_; ++k) {
            P_(u, k) = dist(rng_);
        }
    }

    for (int i = 0; i < n_items_; ++i) {
        for (int k = 0; k < n_factors_; ++k) {
            Q_(i, k) = dist(rng_);
        }
    }

    bu_.setZero();
    bi_.setZero();
}

void MatrixFactorizationSGD::validate_indices(const std::vector<Rating>& ratings) const {
    for (const auto& r : ratings) {
        if (r.user < 0 || r.user >= n_users_) {
            throw std::out_of_range("user index out of range.");
        }
        if (r.item < 0 || r.item >= n_items_) {
            throw std::out_of_range("item index out of range.");
        }
    }
}

void MatrixFactorizationSGD::fit(const std::vector<Rating>& ratings, bool verbose) {
    if (ratings.empty()) {
        throw std::invalid_argument("ratings must not be empty.");
    }

    validate_indices(ratings);

    global_mean_ = 0.0;
    for (const auto& r : ratings) {
        global_mean_ += r.value;
    }
    global_mean_ /= static_cast<double>(ratings.size());

    std::vector<Rating> shuffled = ratings;

    for (int epoch = 0; epoch < n_epochs_; ++epoch) {
        std::shuffle(shuffled.begin(), shuffled.end(), rng_);

        for (const auto& r : shuffled) {
            int u = r.user;
            int i = r.item;
            double y = r.value;

            double pred = global_mean_ + bu_(u) + bi_(i) + P_.row(u).dot(Q_.row(i));
            double err = y - pred;

            Eigen::RowVectorXd pu = P_.row(u);
            Eigen::RowVectorXd qi = Q_.row(i);

            bu_(u) += lr_ * (err - reg_ * bu_(u));
            bi_(i) += lr_ * (err - reg_ * bi_(i));

            P_.row(u) += lr_ * (err * qi - reg_ * pu);
            Q_.row(i) += lr_ * (err * pu - reg_ * qi);
        }

        if (verbose) {
            double rmse = compute_rmse(ratings);
            std::cout << "[Epoch " << (epoch + 1) << "/" << n_epochs_
                      << "] RMSE = " << rmse << std::endl;
        }
    }
}

double MatrixFactorizationSGD::predict(int user, int item) const {
    if (user < 0 || user >= n_users_) {
        throw std::out_of_range("user index out of range.");
    }
    if (item < 0 || item >= n_items_) {
        throw std::out_of_range("item index out of range.");
    }

    return global_mean_ + bu_(user) + bi_(item) + P_.row(user).dot(Q_.row(item));
}

Eigen::MatrixXd MatrixFactorizationSGD::full_prediction() const {
    Eigen::MatrixXd pred =
        Eigen::MatrixXd::Constant(n_users_, n_items_, global_mean_);
    pred.colwise() += bu_;
    pred.rowwise() += bi_.transpose();
    pred += P_ * Q_.transpose();
    return pred;
}

double MatrixFactorizationSGD::compute_rmse(const std::vector<Rating>& ratings) const {
    double sse = 0.0;
    for (const auto& r : ratings) {
        double err = r.value - predict(r.user, r.item);
        sse += err * err;
    }
    return std::sqrt(sse / static_cast<double>(ratings.size()));
}

const Eigen::MatrixXd& MatrixFactorizationSGD::user_factors() const {
    return P_;
}

const Eigen::MatrixXd& MatrixFactorizationSGD::item_factors() const {
    return Q_;
}

const Eigen::VectorXd& MatrixFactorizationSGD::user_bias() const {
    return bu_;
}

const Eigen::VectorXd& MatrixFactorizationSGD::item_bias() const {
    return bi_;
}

double MatrixFactorizationSGD::global_mean() const {
    return global_mean_;
}
\end{lstlisting}

\end{document}